\documentclass[prd,showpacs,letterpaper,twocolumn]{revtex4}%
\usepackage{graphicx}
\usepackage{bm}
\usepackage{epsf}
\usepackage{rotating}
\usepackage{epsfig,graphics,rotate,color}
\usepackage{wrapfig}
\usepackage{amssymb}
\usepackage{amsmath}
\usepackage{amsfonts}
\usepackage{array,hhline,dcolumn}%
\providecommand{\U}[1]{\protect\rule{.1in}{.1in}}
\begin{document}
\title{Limits on Electron Neutrino Disappearance from the KARMEN and 
LSND $\nu_e$-Carbon Cross Section Data}
\author{J.M. Conrad$^{1}$ and M.H. Shaevitz$^{2}$}
\affiliation{$^{1}$ Massachusetts Institute of Technology}
\affiliation{$^{2}$ Columbia University}

\begin{abstract}

  This paper presents a combined analysis of the KARMEN and LSND
  $\nu_e$-carbon cross section measurements within the context of a
  search for $\nu_e$ disappearance at high $\Delta m^2$.  KARMEN and
  LSND were located at 17.7 m and 29.8 m respectively from the
  neutrino source, so, the consistency of the two measurements, as a
  function of antineutrino energy, sets
  strong limits on neutrino oscillations.  Most of the
 allowed region from the $\nu_e$ disappearance analysis of the Gallium calibration data 
  is excluded at $>$95\% CL and the best fit point is
  excluded at 3.6$\sigma$.  Assuming CPT conservation, comparisons
  are also made to the oscillation analyses of reactor
  antineutrino data.
   
\end{abstract}

\pacs{14.60.Pq,14.60.St}
\maketitle

\section{Introduction}

This paper presents an analysis of the $\nu_e$-carbon cross section data from
the KARMEN \cite{KARMEN} and LSND \cite{LSND} experiments, within the
context of electron neutrino oscillations at high $\Delta m^2$.  In a
two-neutrino oscillation formalism, the probability for $\nu_e$
disappearance is given by:
\begin{equation}
P=1- \sin^2
2\theta \ \sin^2(1.27 \Delta m^2(L/E)),
\label{osc}
\end{equation}
where $\theta$ is the mixing angle; $\Delta m^2 =
m_{2}^2-m_{1}^2$, in eV$^2$, is the difference between the squared
neutrino masses; $L$, in m, is the distance from the neutrino source
to the detector; and $E$, in MeV, is the neutrino energy.  This
analysis exploits the fact that KARMEN and LSND were located at
$L=$17.7 m and 29.8 m respectively.  We use the consistency between
the cross section measurements to place strong
constraints on $\nu_e$ disappearance at $\Delta m^2 \sim 1$~eV$^2$.

This study is motivated by recent results 
that can be interpreted as oscillations with $\Delta m^2 \sim
1$~eV$^2$.  The strongest evidence comes from the LSND experiment,
which observed a $\bar \nu_\mu \rightarrow \bar \nu_e$ signal
corresponding to an oscillation probability of
 $(0.264 \pm 0.067 \pm 0.045)\%$ \cite{LSNDosc}. 
Recent MiniBooNE antineutrino data \cite{MBnubar} are
in agreement with LSND, although with less significance, while the
MiniBooNE neutrino data do not support $\nu_\mu \rightarrow \nu_e$
oscillations \cite{MBnu}.

High $\Delta m^2$ muon-to-electron flavor appearance cannot be
explained in a three neutrino mixing model that also incorporates
``solar'' and ``atmospheric'' oscillations \cite{Sorel}.  As a result,
these data have inspired models with three active and one sterile
(3+1) or 2 sterile (3+2) neutrinos.  Sterile neutrinos ($\nu_s$) do
not interact via the weak interaction, but can mix with and cause
oscillations between, the active flavors.  These models
predict a $\nu_e \rightarrow \nu_s$ signal \cite{Viability} with a
large $\Delta m^2$ (on the order of a few eV$^2$) compared to the
splittings between the light states (of order $\sim 10^{-3}$ and $10^{-4}$
eV$^2$).  Therefore, one can take the three light states to be
effectively degenerate.  

This degeneracy simplifies the 3+1 model to an approximate
two-neutrino oscillation model for both appearance and disappearance.
As a result, Eq.~\ref{osc} will be applicable to the following
discussion, where we will use $\theta_{ee}$ as the mixing angle
relevant to $\nu_e$ disappearance.

Recently, a reanalysis of reactor $\bar \nu_e$ flux predictions
\cite{Mueller} has provoked further interest in electron flavor
disappearance in 3+1 models \cite{Mention, Giunti1, Kopp}.  This new analysis resulted
in a shift of the ratio of reactor data-to-prediction from $0.976 \pm
0.024$ to $0.943\pm 0.023$.  This deficit with respect to
prediction is called the ``Reactor Anomaly'' in this paper. This can
be taken as indication of $\bar \nu_e \rightarrow \bar \nu_s$ in a 3+1
model at 98.6\% CL \cite{Mention}.  The best fit is $\Delta m^2 =
1.78$ eV$^2$ and $\sin^2 2\theta_{ee} =0.088$ \cite{Kopp}.

Indications of $\nu_e$ disappearance have arisen from calibration data taken
by the SAGE \cite{SAGE3} and GALLEX \cite{GALLEX3} experiments.  These
used megacurie sources of $^{51}$Cr and $^{37}$Ar to calibrate the
$\nu_e + ^{71}{\rm Ga} \rightarrow ^{71}{\rm Ge} + e^-$ experiments.
The data from SAGE and GALLEX are consistent, and show a
measured-to-predicted ratio of $R=0.86\pm0.05$ \cite{Giunti1}.  We
refer to this as the ``Gallium data'' in this paper. This can be
interpreted as a 2.7$\sigma$ indication of $\nu_e \rightarrow \nu_s$
oscillations \cite{Giunti1, Giunti2}.  The best fit in a 3+1 model
corresponds to a $\Delta m^2 = 2.24$ eV$^2$ and
$\sin^22\theta_{ee}=0.50$ \cite{Giunti1}.  This 
apparent $\nu_e$ disappearance signal
leads to the argument \cite{GiuntiMB} that
$\nu_e \rightarrow \nu_s$,
consistent with the Gallium data, must be applied to the intrinsic
$\nu_e$ background of the MiniBooNE $\nu_\mu \rightarrow \nu_e$
search \cite{MBnu}.

\begin{figure}[b]\begin{center}
{\includegraphics[width=2.75in]{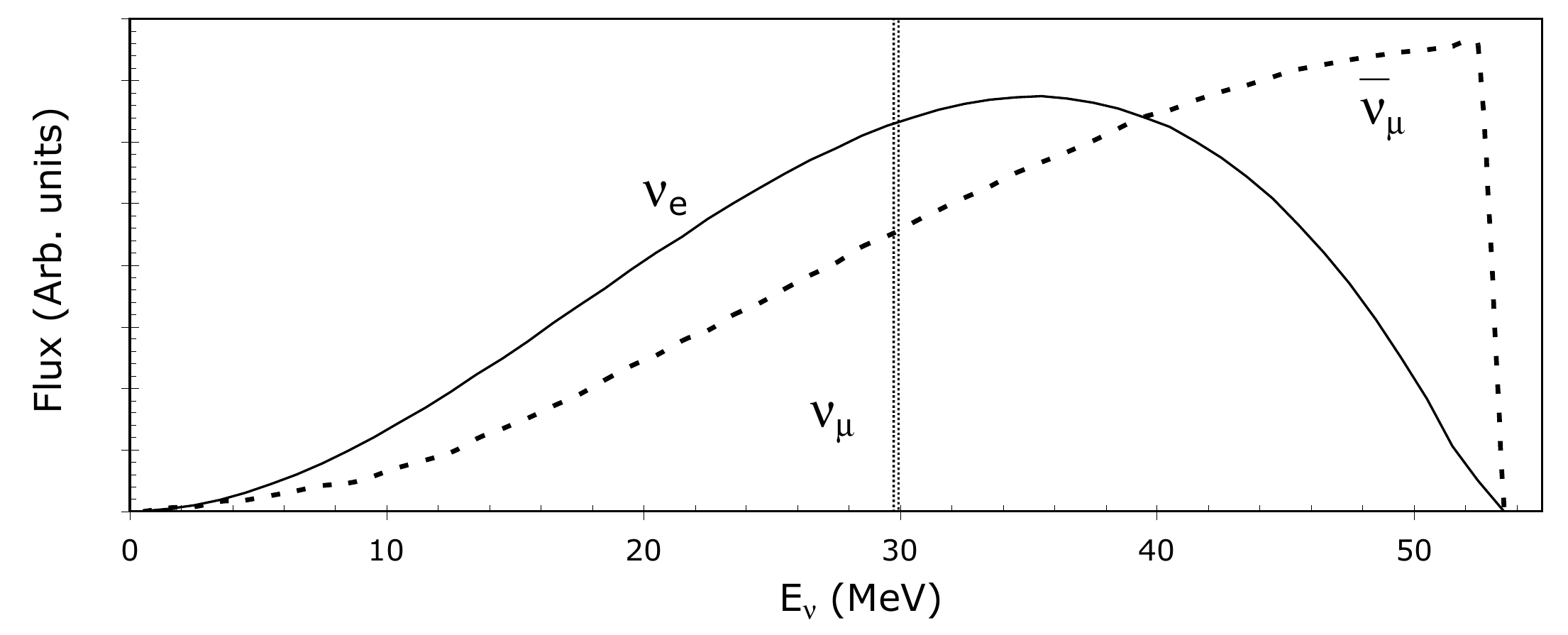}
} 
\end{center}
\vspace{-0.25in}
\caption{Energy distribution of neutrinos in a DAR beam.
\label{flux} }
\end{figure}

The results of the 3+1 models have provoked
substantial interest within the community,  but they are not
decisive.
As a result they are a prime
motivation for further studies of electron neutrino disappearance
\cite{LENANOvA, Palazzo}.  There are few opportunities for precision $\nu_e$
disappearance searches, since most beams have large uncertainties in
the normalization and energy distribution for the $\nu_e$ and $\bar
\nu_e$ beam contents.  However, decay-at-rest (DAR) neutrino beams
can provide a unique window on electron-flavor neutrino oscillations.

\section{KARMEN and LSND}

KARMEN and LSND were DAR experiments that ran in the 1990's using 800
MeV protons on target.  The isotropic DAR flux, shown in
Fig.~\ref{flux}, has equal $\nu_\mu$, $\bar \nu_\mu$ and $\nu_e$
content with a well-understood energy spectrum described by weak
decay physics.  The beam energy extends to 52.8 MeV.  The
normalization is known to 10\%, with the uncertainty dominated by the pion
production rate per incident proton \cite{BurmanISIS,BurmanLAMPF}.

KARMEN ran at the ISIS facility at Rutherford Laboratory, with
200 $\mu$A of protons impinging on a copper, tantalum, or uranium
target.  The center of the nearly cubic detector was located at 17.7 m
from the proton target, at an angle of 100$^\circ$.  The liquid
scintillator target volume was 56 m$^3$ and consisted of 512 optically
independent modules (17.4 cm $\times$ 17.8 cm $\times$ 353 cm) wrapped
in Gadolinium-doped paper.  More details are available in
Ref.~\cite{Kardet}.

LSND used protons from the LAMPF accelerator at Los Alamos National
Laboratory (LANL), where a 1 mA
beam of protons impinged on a water target.  The center of the
8.75 m long, nearly cylindrical detector was located at 29.8 m from
the target, at an angle of 12$^\circ$ from the proton beam direction.
This was an unsegmented detector with a fiducial mass of 167 tons of
oil (CH$_2$), lightly doped with b-PBD scintillator.  
More details are available in Ref.~\cite{LSNDdet}.

Both experiments measured $\nu_e +^{12}{\rm C} \rightarrow ^{12}{\rm
  N}_{gs} + e^-$ scattering.  In this two-body interaction, with
$Q$-value of 17.3 MeV, the neutrino energy can be reconstructed by
measuring the outgoing visible energy of the electron.  The $^{12}$N
ground state is identified by the subsequent $\beta$ decay,
$^{12}{\rm N}_{gs} \rightarrow ^{12}{\rm C} + e^+ + \nu_e$, which has
a $Q$-value of 16.3 MeV and a lifetime of 15.9 ms.

\section{The KARMEN and LSND Cross Sections \label{xsecintro}}

\begin{figure}[t]\begin{center}
{
\includegraphics[width=2.95in]{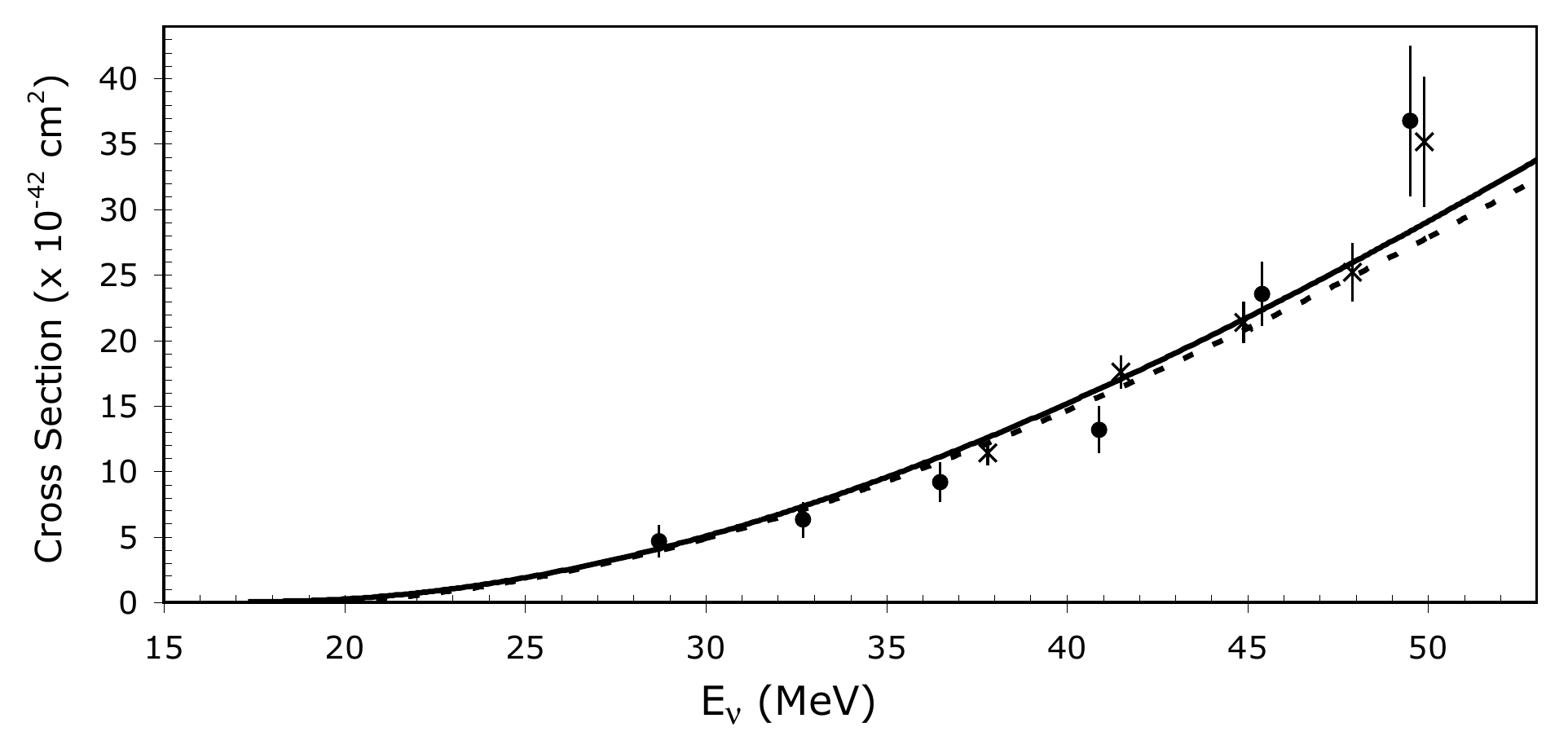}
} \end{center}
\vspace{-0.25in}
\caption{The KARMEN (points) and LSND (crosses) 
measured cross sections with statistical errors for 
$\nu_e +^{12}{\rm C}
\rightarrow ^{12}{\rm N}_{gs} + e^-$ compared to the theoretical prediction
of Fukugita, {\it et al.} (solid line), based on the EPT model, 
and Kolbe, {\it et al.} (dashed line), based on the CRPA model. 
\label{xsecs} }
\end{figure}

The KARMEN and LSND cross section measurements for $\nu_e +^{12}{\rm
  C} \rightarrow ^{12}{\rm N}_{gs} + e^-$ \cite{KARMEN,LSND}, in
energy bins, are compared in Fig.~\ref{xsecs}.  The corresponding
flux-averaged cross sections measured by KARMEN and LSND are given in
Table~\ref{fluxavetab}.  For completeness, we also list the
flux-averaged cross section for the LANL E225 experiment \cite{Chen},
which was located 9 m from a DAR source.  (E225 did not publish
energy-binned cross section measurements.)  
The agreement between all three experiments is
excellent.

 The measured cross sections are compared to predictions by Fukugita,
{\it et al.}\cite{Fukugita} and by Kolbe {\it et al.}\cite{Kolbe} in
Fig.~\ref{xsecs}.  Both models follow an $(E_\nu-Q)^2$ form, where
$Q=17.3$ MeV.  This energy dependence arises because the interaction
is an allowed transition, converting the $0^+$ ($^{12}$C) state to the
$1^+$ ($^{12}$N) state. The Fukugita prediction is calculated within
the ``elementary particle model'' (EPT) and has an associated 12\%
normalization uncertainty \cite{Fukugita}.  Other EPT predictions,
include Donnelly \cite{Donnelly} and Mintz, {\it et al.}, are given in
Table~\ref{fluxavetab}.  For comparison, the Kolbe, {\it et al.},
calculation \cite{Kolbe} is performed within a ``continuum random
phase approximation'' (CRPA) approach.  A discussion of the relative
merits of EPT versus CRPA models for describing this process
appears in Ref.~\cite{Vogel}.
From a strictly experimental point of view, both EPT and CRPA models
fit the data well.  To be clear, these theoretical results are true
predictions rather than fits, since they were published well before
the KARMEN and LSND results.

\begin{table}[tbp] \centering
\begin{tabular}
[c]{l|c|c}%
\hline
Experiment (Dist.)& Flux-Averaged Cross Section & Ref. \\ \hline
KARMEN (17.7 m)& $(9.1\pm 0.5\pm 0.8) \times 10^{-42}$ cm$^2$ & \cite{KARMEN} \\
LSND (29.8 m) & $(8.9\pm 0.3 \pm 0.9) \times 10^{-42}$ cm$^2$ & \cite{LSND} \\
E225 (9 .0 m)& $(1.05 \pm 0.10 \pm 0.10) \times 10^{-41}$ cm$^2$ & \cite{Chen} \\ \hline\hline
Prediction & Flux-Averaged Cross Section & Ref. \\ \hline
Fukugita {\it et al.} & $9.2 \times 10^{-42}$ cm$^2$ & \cite{Fukugita}  \\
Mintz {\it et al.}&  $8.0 \times 10^{-42}$ cm$^2$ & \cite{Mintz} \\
Donnelly & $9.4 \times 10^{-42}$ cm$^2$  & \cite{Donnelly} \\
Kolbe {\it et al.} &  $8.9 \times 10^{-42}$ cm$^2$ & \cite{Kolbe} \\ \hline
\end{tabular}
\caption{Top: Flux-averaged $\nu_e + ^{12}{\rm C} \rightarrow e^+ +
^{12}{\rm N}_{gs}$ cross section measurements 
with statistical and systematic error. 
  Bottom:  Flux-averaged predictions from EPT (Fukugita, Mintz and Donnelly) 
  and CRPA (Kolbe) models. Flux-average cross section values are equivalent 
  to those for a neutrino of 35 MeV energy.}
  \label{fluxavetab}
\end{table}

\section{Constraints on Electron Neutrino Disappearance}

The allowed regions for $\nu_e \rightarrow \nu_s$ oscillations are
determined from a comparison of the LSND and KARMEN data with respect
to the Fukugita prediction.  For a given oscillation hypothesis
($\Delta m^2$ and $\sin^2 2\theta_{ee}$), we calculate a combined $\chi^2$
for LSND and KARMEN with respect to the prediction using the statistical
error for each data point and employing three pull terms as a method
to incorporate systematic uncertainties.  The first pull term
represents the correlated normalization error.  As noted in the KARMEN paper
\cite{KARMEN}, LSND and KARMEN have a 7\% systematic error on the
neutrino flux normalization from the flux simulation that is
correlated between the two experiments \cite{BurmanISIS, BurmanLAMPF}.
This is combined in quadrature with the 12\% systematic error on the
normalization for the Fukugita prediction to give the correlated
normalization pull term in the $\chi^2$ calculation.  The remaining
uncorrelated normalization uncertainties for each experiment are 7\%
for LSND \cite{LSND} and 5\%  for KARMEN \cite{KARMEN}.
These uncertainties are used as the two other pull terms in the
$\chi^2$ calculation. To determine the 90\% CL allowed regions in
$\Delta m^2$ and $\sin^2 2\theta_{ee}$, we marginalize over the three
normalization pull parameters and use a $\Delta \chi^2 > 4.61$ 
requirement for the two-degrees-of-freedom excluded region.

\begin{figure}[t]\begin{center}
{
\includegraphics[width=2.75in]{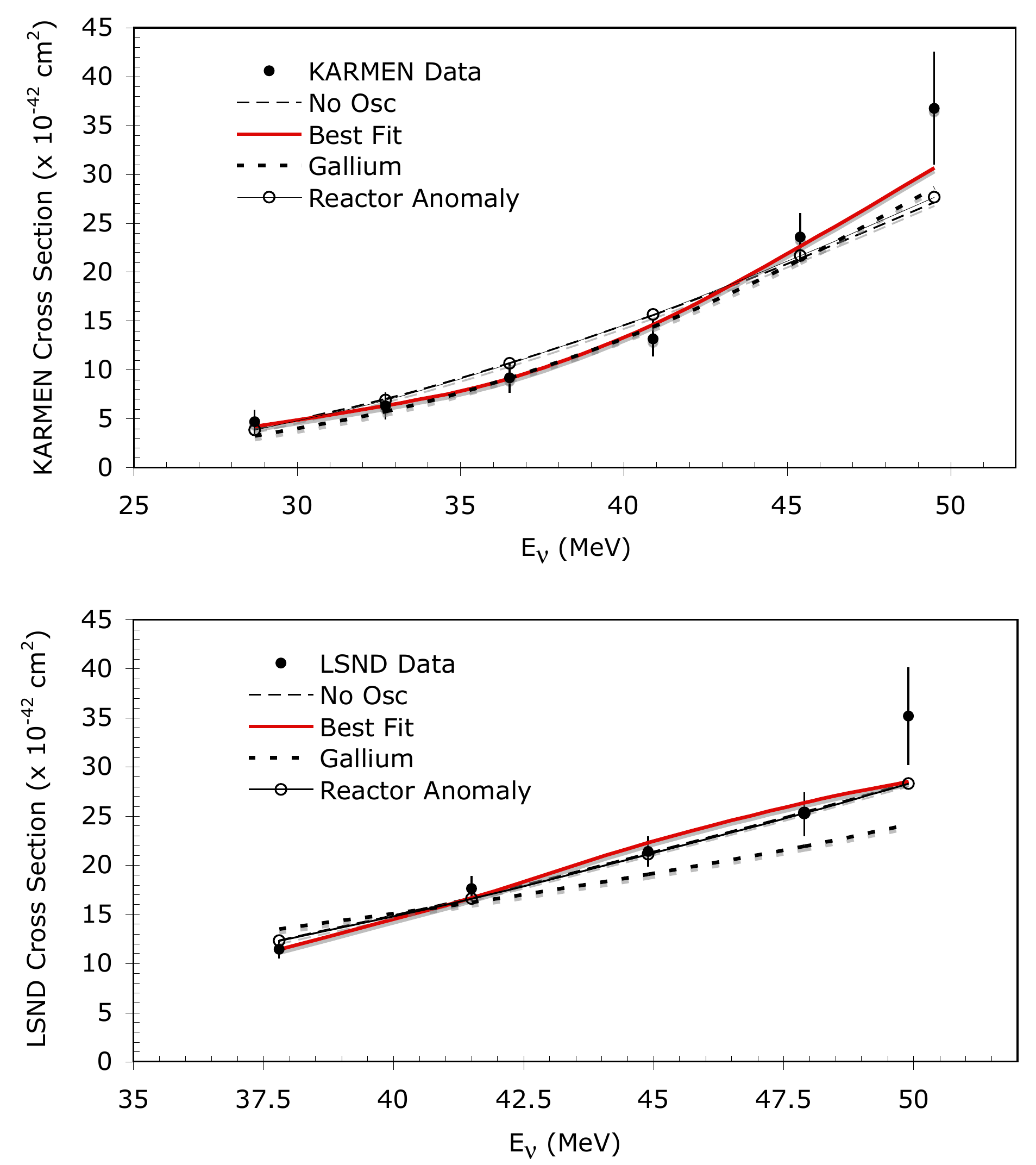}
} \end{center}
\vspace{-0.25in}
\caption{Comparisons of the data to various oscillation predictions
for the LSND (top) and KARMEN (bottom) data 
using the Fukugita prediction, as described in the text.
\label{fits} }
\end{figure}


\begin{table}[tbp] \centering

\begin{tabular}
[c]{lcrr}%
\multicolumn{2}{l}{Fukugita et al. Cross Section} &  & \\\hline
Fit Type &  $\Delta m^2$(eV$^2$) & \multicolumn{1}{c}{$\sin^2 2\theta_{ee}$} &
\multicolumn{1}{c}{$\chi^2$/DOF}\\\hline
Best Fit & 7.49 $\pm$ 0.39 & \multicolumn{1}{c}{0.290 $\pm $0.115} &
\multicolumn{1}{c}{5.5/9}\\
No Osc & --- & \multicolumn{1}{c}{0.000 (fixed)} & \multicolumn{1}{c}{10.4/11}\\
Gallium & 2.24 (fixed) & \multicolumn{1}{c}{0.500 (fixed)} &
\multicolumn{1}{c}{34.3/11}\\
Reactor Anomaly & 1.78 (fixed) & \multicolumn{1}{c}{0.089 (fixed)} &
\multicolumn{1}{c}{10.2/11}\\\hline
\multicolumn{1}{r}{} & \multicolumn{1}{r}{} &  & \\

\multicolumn{2}{l}{Kolbe et al. Cross Section} &  & \\\hline
Fit Type &  $\Delta m^2$(eV$^2$) & \multicolumn{1}{c}{$\sin^2 2\theta_{ee}$} &
\multicolumn{1}{c}{$\chi^2$/DOF}\\\hline
Best Fit & 7.49 $\pm$ 0.39 & \multicolumn{1}{c}{0.281 $\pm$ 0.115} &
\multicolumn{1}{c}{6.1/9}\\
No Osc & --- & \multicolumn{1}{c}{0.000 (fixed)} & \multicolumn{1}{c}{10.7/11}\\
Gallium & 2.24 (fixed) & \multicolumn{1}{c}{0.500 (fixed)} &
\multicolumn{1}{c}{37.8/11}\\
Reactor Anomaly & 1.78 (fixed) & \multicolumn{1}{c}{0.089 (fixed)} &
\multicolumn{1}{c}{10.8/11}\\\hline
\end{tabular}
\caption{Results of fits using the Fukugita (top) or Kolbe (bottom) 
cross sections for the predicted energy dependence.  The resultant
or assumed $\Delta m^2$ and $\sin^2 2\theta_{ee}$ values, along with the $\chi^2$ and degrees
of freedom (DOF) for the fits are shown.}\label{chi2}
\end{table}

\begin{figure}[t]\begin{center}
{
\includegraphics[width=2.95in]{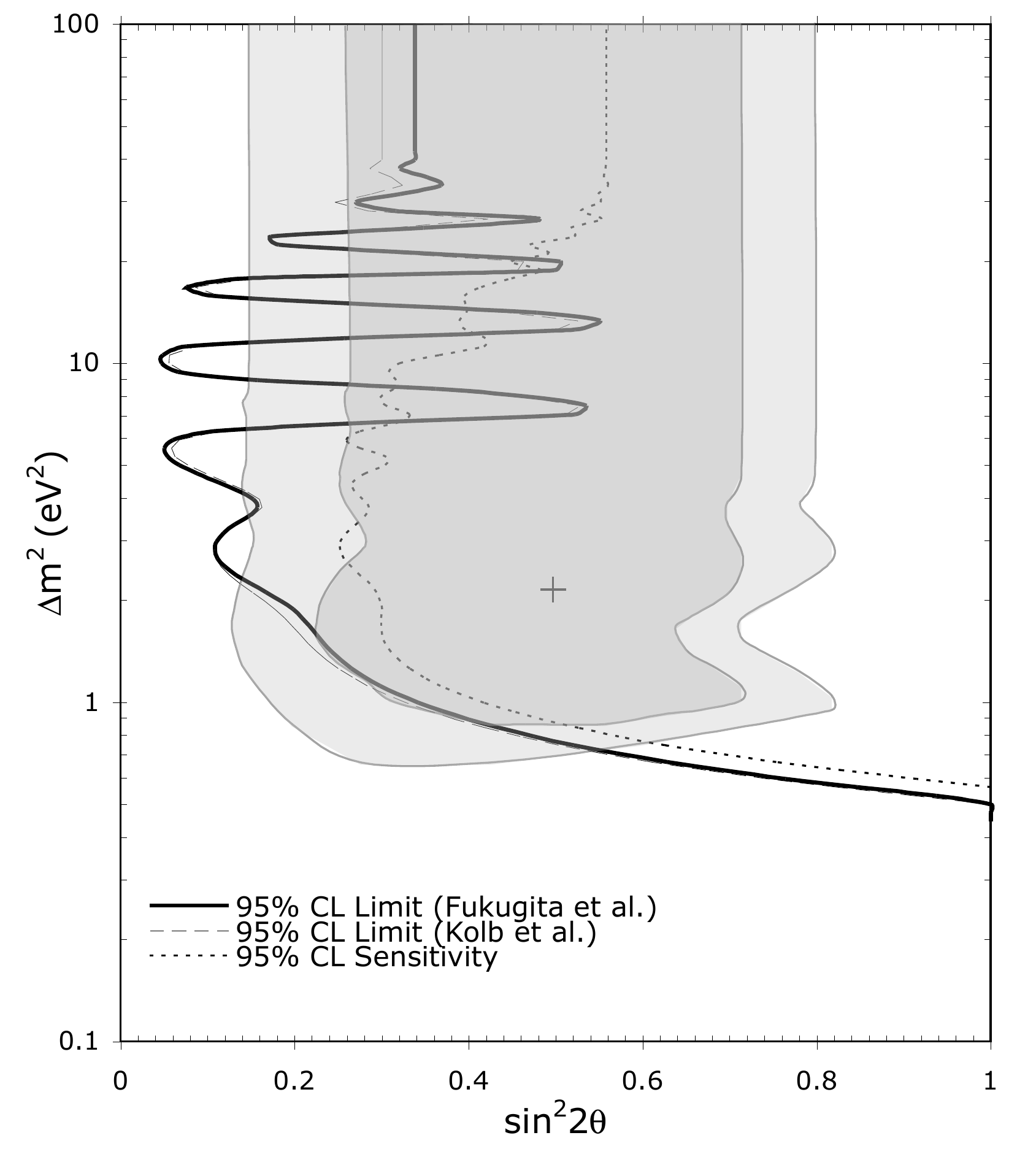}\\
} \end{center}
\vspace{-0.25in}
\caption{The 95\% $\nu_e$ disappearance limit from the Fukugita (EPT) fit
  (solid, black line) compared to the predicted sensitivity
  (dotted line).  Also shown is the 68\% (darker, shaded region) and
  90\% (lighter, shaded region) contours from the Gallium experiments.  The
  dashed line is the Kolbe (CRPA) fit.  
  \label{contour} }
\end{figure}

The results of the fits using the Fukugita prediction are shown in
Fig.~\ref{fits}.  Table~\ref{chi2} reports the $\chi^2$ and degrees of
freedom (DOF) for various joint fits to the LSND and
KARMEN data points.  The fit without oscillations (No Osc), shown as the
long-dashed line in Fig.~\ref{fits}, has a $\Delta \chi^2$ 
probability of 91.5\% and is only excluded at the 1.7$\sigma$ level. As a result, we use the data to set a limit on $\nu_e$
disappearance and calculate the 95\% CL exclusion
region  shown in Fig.~\ref{contour}.  The
best fit, indicated by the solid lines in Fig.~\ref{fits}, is at
$\Delta m^2 = 7.49 \pm 0.39$ eV$^2$ and $\sin^2 2\theta_{ee} = 0.290 \pm 0.115$.

Comparing the data to an oscillation model with the best-fit Gallium parameters 
illustrates the strong disagreement, though we note that the Gallium fit had
a rather shallow minimum \cite{Giunti1, Giunti2}. 
The Gallium fit reported in Table~\ref{chi2} and shown as the dashed
line on Fig.~\ref{fits} is poor.  This point
has a $\chi^2$ 
probability of less than $3.2 \times 10^{-4}$ and is, therefore, ruled out
at $3.6\sigma$.   (The $\Delta \chi^2$ for this point has a probability of 
$5.3 \times 10^{-7}$, which corresponds to a $5.0\sigma$ exclusion.)
Most of the Gallium allowed region, indicated at 68\% 
and 90\% CL on Fig.~\ref{contour}, is excluded at 95\% CL by
this analysis.

As discussed above, all models tend to follow a
$(E_\nu-Q)^2$ dependence.  Nevertheless, small differences between the
Fukugita (EPT) and Kolbe (CRPA) predictions, shown in Fig.~\ref{xsecs}, allow a
test for model dependence.  The Kolbe fit proceeds in the same way as
for the Fukugita model.  The resulting $\chi^2$ values for the
fits are given in Table~\ref{chi2}.  The comparisons of the fits with
the data are indistinguishable from those shown in Fig.~\ref{fits} and
so are not shown here.  This leads to the conclusion that there is no
substantial systematic effect from the energy dependence of the
underlying cross section model.  The 95\% CL exclusion limit
from the Kolbe fit is also shown in
Fig.~\ref{contour} as the dashed contour and is very similar to the
Fukugita contour.

The excluded region in Fig.~\ref{contour} is better than the
expected sensitivity region (dotted contour)
calculated for an underlying null oscillation hypothesis.
As a way of quantifying this difference, using the Fukugita model fit,
the Gallium data point is
ruled out in a $\chi^2$ analysis at 3.6$\sigma$, while the sensitivity would
have predicted that, for an average experiment with no signal, the
Gallium point would be ruled out at 2.8$\sigma$.  This strong limit
with respect to the sensitivity is not unlikely; we find that 11\% of
simulated experiments have a high $\Delta m^2$ 95\% CL limit for $\sin^2
2\theta_{ee}$ at or below the 0.34 95\% CL limit of this analysis.  
These studies also show that $\Delta \chi^2$ is
a good statistic for determining the exclusion regions since 10\% of
the simulated experiments have a $\Delta \chi^2$ value for the null
oscillation hypothesis greater than 4.61 as expected for two degrees
of freedom.

\begin{figure}[t]\begin{center}
{
\includegraphics[width=2.95in]{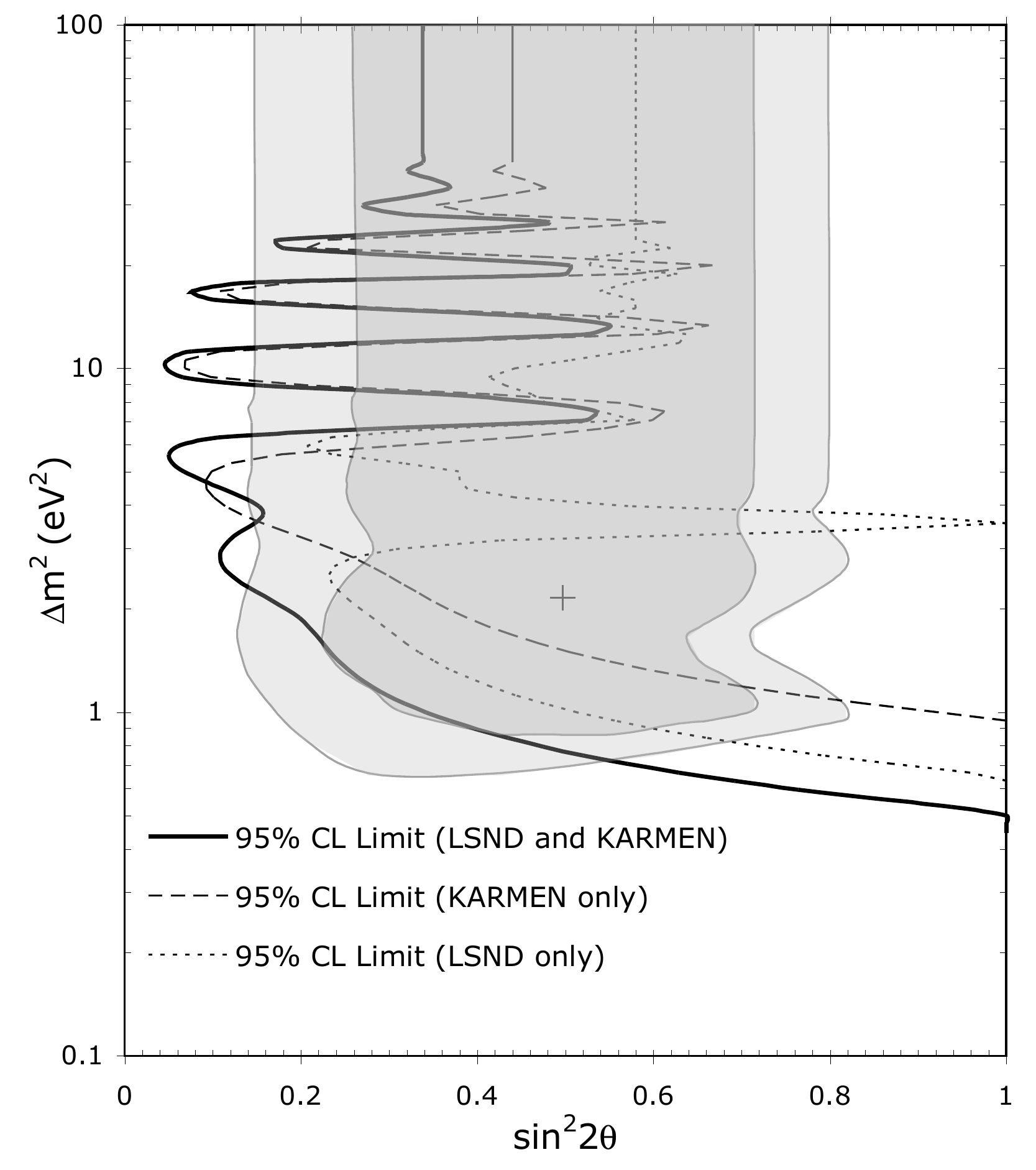}\\
} \end{center}
\vspace{-0.25in}
\caption{The 95\% $\nu_e$ disappearance limit from the combined fit
  (solid, black line) compared to individual fits to KARMEN data (dashed)
  and LSND data (dotted).    This is overlaid on the Gallium 90\% and
  68\% CL allowed regions.   All fits use the Fukugita model.
  \label{individual} }
\end{figure}

The combined fit to KARMEN and LSND is stronger than fits to the 
the individual data sets because the detectors are at different
distances.  Requiring the proper $L$ as well as $E$ dependence
adds an important constraint to the oscillation
fit.    The results of the individual fits are compared to the
combined fit in Fig.~\ref{individual}.  
A fit to only the KARMEN data yields a 
best fit $\chi^2$ of 2.46 for 4 degrees of freedom, with the 
parameters  $\sin^2 2\theta = 0.333 \pm 0.130$ and $\Delta m^2 = 7.54$
eV$^2$.    The $\chi^2$ for the null fit was 7.05 for 6 degrees of 
freedom.   A fit to only the LSND data results in a 
best fit $\chi^2$ of 2.27 for 3 degrees of freedom, with the 
parameters  $\sin^2 2\theta = 0.209 \pm 0.331$ and $\Delta m^2 = 3.90$
eV$^2$.    The $\chi^2$ for the null fit was 3.29 for 5 degrees of 
freedom.     

Considering fits to these two data sets separately allows interpretation of 
certain features in the combined fit.   We see in Fig.~\ref{individual} that 
the KARMEN data dominates at
high $\Delta m^2$ because of statistics and that   
the large variations in the limit are driven by this data set.  These 
variations appear because the event energy distribution
spans a limited range.  As a result,  there are oscillation parameters for which the KARMEN
distance allows the experiment to either be very sensitive or very insensitive to
disappearance.  
The LSND data set allows full
oscillation
at $\Delta m^2$ of 3.9 eV$^2$, but KARMEN data does not.
At low $\Delta m^2$, the sensitivity is dominated by the LSND 
data although the combination with KARMEN is significantly better.


\section{Broader Interpretation of this Constraint}

The limit presented here and the Gallium data represent the only 
electron neutrino disappearance results in this $\Delta m^2$ range.
Comparison to other data sets require interpretation within models.
Specific global analyses are beyond the scope of this paper,  however,
we can consider the impact of the limit, in general.

CPT conservation requires that $\nu_e$ and $\bar \nu_e$ disappearance 
should occur at the same rate.    Because CPT conservation is embedded
in all field theories, comparison of electron flavor neutrino and
antineutrino disappearance is widely regarded as interesting.
The relavent antineutrino disappearance data comes from the reactor
experiments.    The recently published 
Reactor Anomaly hints at oscillations, and the parameters 
from the $\bar \nu_e$ disappearance data \cite{Mention} 
can be compared to the
$\nu_e$ results assuming  CPT conserving models.    

\begin{figure}[t]\begin{center}
{
\includegraphics[width=2.95in]{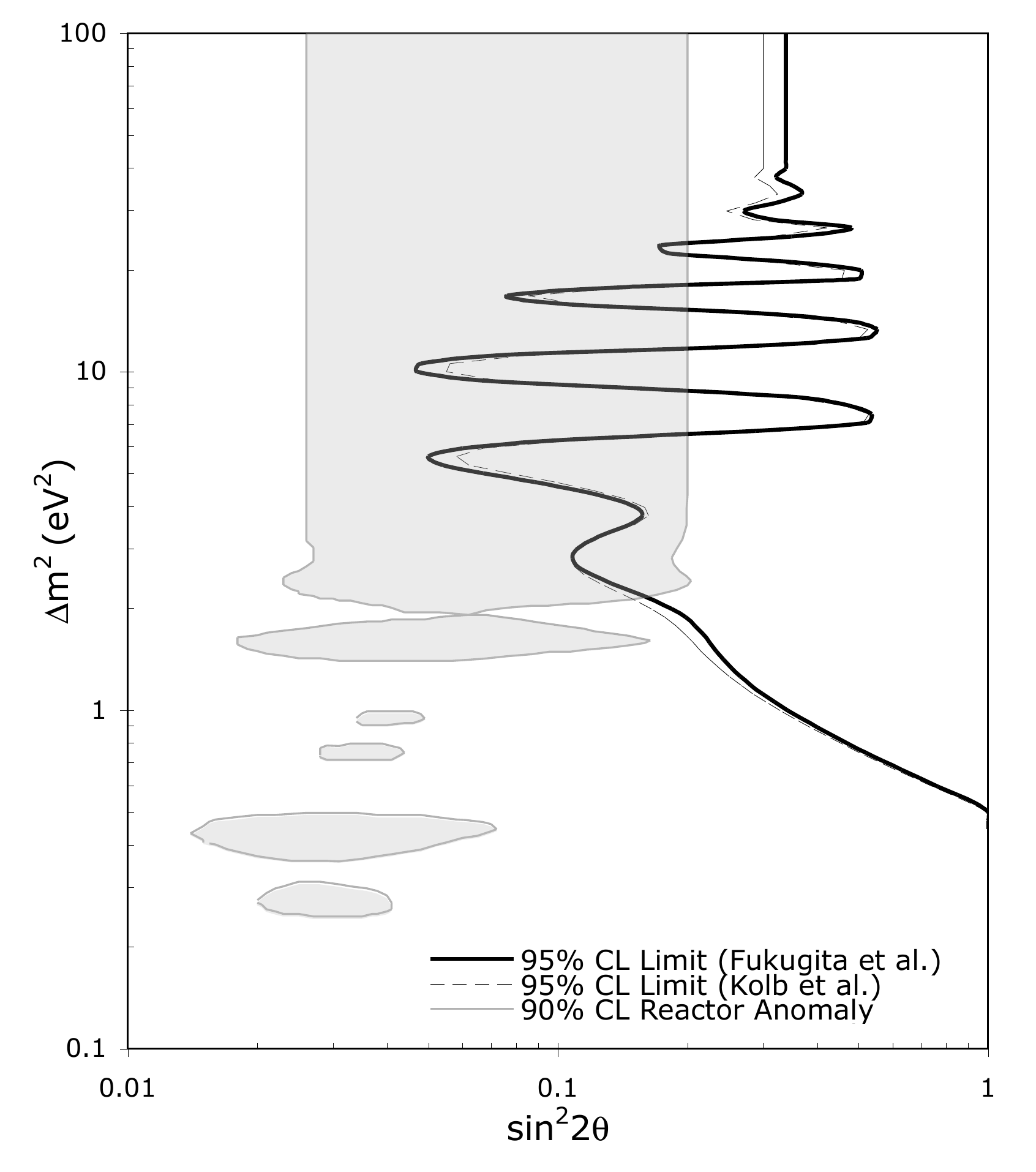}\\
} \end{center}
\vspace{-0.25in}
\caption{The 95\% $\nu_e$ disappearance limit from the Fukugita (EPT) fit
  (solid, black line)  and Kolbe (CRPA) fit compared to the
  90\% (shaded region) contours from the Reactor Anomaly.   
  \label{reactcontour} }
\end{figure}

In Fig.~\ref{reactcontour}, we overlay the Reactor Anomaly allowed 
region at 68\% and 90\% CL with
the 95\% CL limit from this paper using Fukugita.  In this figure,
we use a log scale for $\sin^2 2\theta$ so that the reactor
allowed region is clear.
One can see that portions of the allowed reactor space will be  
excluded in fits that require CPT conservation. 
However, the Reactor Anomaly best fit parameters
give a  $\Delta \chi^2$ probability of 90.8\%, which
lies outside of the 95\% CL excluded region from the cross
section analysis limit, and one expects portions of the allowed
region to survive in a global fit.   The
comparison of the best fit with data, shown as the thin line with dots in
Fig.~\ref{fits}, is reasonably good and similar to the
``no oscillation" case, illustrating why the KARMEN/LSND data does not speak
to the entire region of the Reactor Anomaly.    This is consistent
with good the $\chi^2$/DOF for the best fit point of the reactor
anomaly, reported in  Table~\ref{chi2}.

Comparisons to electron-to-muon flavor appearance data is much more
model dependent than the comparison of $\bar \nu_e$ to $\nu_e$
disappearance.  In this case, one must go beyond choosing whether or
not to assume CPT conservation, to include the number of sterile
neutrinos (typically, one assumes 3+1 or 3+2 models); the mass
heirarchy (a 1+3+1 model has recently been published \cite{Kopp} as a
variation of the two sterile neutrino case); and whether or not to
include CP violation.  

Inclusion of the appearance data also,
necessarily, requires reference to muon-flavor disappearance
as well as electron flavor disappearance, since these are all
realted.   As an example, in a 3+1 model:
\begin{eqnarray}
\sin^22\theta_{\mu e} & = & 4 U_{e4}^2U_{\mu 4}^2,  \label{lsndeq}\\
\sin^22\theta_{\mu \mu} & = & 4 U_{\mu 4}^2 (1-U_{\mu 4}^2), \label{mudiseq} \\
\sin^22\theta_{e e} & = & 4 U_{e 4}^2 (1-U_{e 4}^2);  \label{ediseq}
\end{eqnarray}
where $U_{\alpha i}$ represents the element in the 4-neutrino mixing
matrix.  The disappearance limit we present here, when interpreted
within Eq.~\ref{ediseq}, substantially limits the range of $U_{e4}^2$,
and will considerably reduce the space of allowed values for matrix
elements that can describe muon-to-electron flavor appearance in
neutrino mode, as may be implied by the MiniBooNE low energy excess \cite{MBnu}.  
While the correspondences between experimentally measured mixing
angles and the underlying matrix elements for two-sterile-neutrino
models is more complicated, the basic point still 
applies.

\section{Conclusions}

This analysis has used the $\nu_e +^{12}{\rm C} \rightarrow ^{12}{\rm N}_{gs} +
e^-$ cross section data from LSND and KARMEN to constrain the amount of $\nu_e$
disappearance oscillations at high $\Delta m^2$.  The good agreement between the
data sets and with the theory, despite different
distances of detectors from the source, leads to an 95\% CL exclusion
region which extends down to $\sin^2 2\theta_{ee} = 0.34$ at high $\Delta m^2$
and to considerably lower values for some $\Delta m^2$ ranges.
Comparison to another underlying cross section model
does not significantly change the excluded region.

The data are in disagreement with the only other $\nu_e$ disappearance
data set in this $\Delta m^2$ region, which comes from the Gallium
experiments.  Large portions of the allowed region for 
$\nu_e$ disappearance analysis of the Gallium calibration data are
ruled out at $>$95\% CL.   As a benchmark, 
the best fit point for Gallium is excluded at
3.6$\sigma$.  This new limit also severely restricts models
addressing the MiniBooNE results, such as Ref.~\cite{GiuntiMB}, 
that incorporate $\nu_e$ disappearance to explain the observed
energy distribution. 

Assuming CPT conservation, this data set can be compared to the $\bar
\nu_e$ disappearance from reactor data.  The Reactor Anomaly best fit point in
this case is within the allowed 95\% CL contour
but some regions of the allowed region in Ref.~\cite{Mention} are excluded.

~~~\\

\begin{center}
{ \textbf{Acknowledgments}}
\end{center}

The authors thank the National
Science Foundation for support.

\end{document}